\documentclass{amsart}
\usepackage{graphicx,amssymb}
\newtheorem{thm}{Theorem}[section]

\theoremstyle{definition}
\newtheorem{defn}[thm]{Definition}
\theoremstyle{remark}

\def\beq{\begin{eqnarray}}
\def\eeq{\end{eqnarray}}
\def\bsp{\begin{split}}
\def\esp{\end{split}}

\def\Tr{\mathrm{Tr}}
\def\d{\mathrm{d}}
\def\diag{\mathrm{diag}}
\def\RC{$CSI_R$}
\def\KC{$CSI_K$}
\def\FC{$CSI_F$}

\newcommand{\mbold}[1]{\mbox{\boldmath{\ensuremath{#1}}}}

\begin{document}

\title[\textbf{Lorentzian CSI spacetimes  in 3D}]{\textbf{Lorentzian spacetimes with constant curvature invariants in three dimensions}}
\author{\textbf{Alan Coley, Sigbj\o rn Hervik, Nicos Pelavas}}
\address{Department of Mathematics and Statistics,
Dalhousie University, Halifax, Nova Scotia, Canada B3H 3J5}
\email{aac@mathstat.dal.ca, herviks@mathstat.dal.ca,
pelavas@mathstat.dal.ca}
\date{\today}
\maketitle

\begin{abstract}

In this paper we study
Lorentzian spacetimes for which all polynomial scalar invariants
constructed from the Riemann tensor and its covariant derivatives
are constant ($CSI$ spacetimes) in three dimensions.
We determine all such $CSI$ metrics explicitly, and show that
for every $CSI$ with particular constant invariants
there is a locally homogeneous spacetime with precisely the same
constant invariants.
We prove that a three-dimensional $CSI$ spacetime
is either (i) locally homogeneous or (ii) it is  locally a
Kundt spacetime.  Moreover, we show that there exists a null frame in which the Riemann (Ricci)
tensor and its derivatives are of boost order zero with
constant boost weight zero components at each order.  Lastly, these
spacetimes can be explicitly constructed from locally homogeneous
spacetimes and vanishing scalar invariant spacetimes.

\vspace{0.3cm}


\end{abstract}


\section{Introduction}

\noindent Lorentzian spacetimes for which all polynomial scalar invariants
constructed from the Riemann tensor and its covariant derivatives
are constant are called $CSI$ spacetimes \cite{CSI}. For a Riemannian
manifold every $CSI$ space is locally homogeneous $(CSI \equiv
H)$ \cite{prufer}. This is not true for Lorentzian manifolds. However, for every
$CSI$ with particular constant invariants there is a homogeneous
spacetime with the same constant invariants.  This suggests that
$CSI$ spacetimes can be constructed from $H$ and vanishing scalar invariants
($VSI$) spacetimes \cite{Higher,4DVSI}. In particular, the relationship
between the various classes of $CSI$ spacetimes (${CSI_R}$,
${CSI_F}$, ${CSI_K}$, defined below) and especially with $CSI
\backslash H$ have been studied in arbitrary dimensions \cite{CSI}.

$CSI$ spacetimes were first studied in \cite{CSI}. It was argued
that for $CSI$ spacetimes that are not locally homogeneous, the
Riemann type is $II$, $D$, $III$, $N$ or $O$ \cite{class}, and that
all boost weight zero terms are constant. The four-dimensional
case was considered in detail, and a number of results and ($CSI$)
conjectures were presented. In \cite{CFH} a number of higher
dimensional $CSI$ spacetimes were constructed that are solutions of
supergravity, and their supersymmetry properties were briefly
discussed.

In this paper we shall study $CSI$ spacetimes in three dimensions
(3D).  Our motivation arises from its relevance to the corresponding study of $CSI$ spacetimes in four and higher dimensions.  There are also applications to solutions of 3D gravity \cite{obukhov}. In particular, we shall determine all 3D $CSI$ metrics, and
prove that if the metric is not locally homogeneous then it is of
Kundt form and all boost weight zero terms are constant, and show
that, by explicitly finding all spacetimes, the 3D $CSI$
spacetimes can be constructed from locally homogeneous spacetimes
and $VSI$ spacetimes.


In the analysis we shall use the canonical Segre forms for the
Ricci tensor in a frame in which the Ricci components are constant
\cite{Milson}. We shall explicitly use these canonical forms to prove a
number of results in 3D \cite{Hall}. We classify different
cases in terms of  their Segre type since the Segre type is more
refined than the Ricci type and in 3D the Riemann tensor is
completely determined by the Ricci tensor. As usual, all of the
results are proven locally in an open neighborhood $U$ (in which
there exists a well-defined canonical frame for which the Segre
type does not change). Hence the results apply in neighborhoods
of all points except for a set of measure zero (these points
typically correspond to boundary points where the algebraic type
such as, for example, the Segre type, can change).
We note that this may result in an incomplete
spacetime (even local homogeneity is no guarantee for completeness).
All possible
$CSI$ spacetimes are then obtained by matching solutions across
boundaries subject to appropriate differentiability conditions
(e.g., we could seek maximal analytic extensions).

\subsection{Preliminaries}

Consider a spacetime $\mathcal{M}$ equipped with a metric $g$. Let us denote by $\mathcal{I}_k$ the set of all scalar
invariants constructed from the curvature tensor and its covariant derivatives up
to order $k$.
\begin{defn}[${VSI}_k$ spacetimes]
$\mathcal{M}$ is called ${VSI}_k$ if for any invariant $I\in\mathcal{I}_k$, $I=0$ over $\mathcal{M}$.
\end{defn}
\begin{defn}[{$CSI$}$_k$ spacetimes]
$\mathcal{M}$ is called {$CSI$}$_k$ if for any invariant $I\in\mathcal{I}_k$, $\partial_{\mu}{I}=0$ over $\mathcal{M}$.
\end{defn}
Moreover, if a spacetime is {$VSI$}$_k$ or {$CSI$}$_k$ for all
$k$, we will simply call the spacetime {$VSI$} or {$CSI$},
respectively. The set of all locally homogeneous spacetimes, denoted by $H$, are the spacetimes for which there exists, in any neighborhood, an isometry group acting transitively. Clearly ${VSI} \subset {CSI}$ and $H \subset{CSI}$.

\begin{defn}[{\RC} spacetimes]
Let us denote by \RC ~all reducible $CSI$ spacetimes that can be
built from $VSI$ and $H$ by (i) warped products (ii) fibered
products, and (iii) tensor sums.
\end{defn}

\begin{defn}[{\FC} spacetimes]
Let us denote by \FC ~those spacetimes for which there exists a
frame with a null vector $\ell$ such that all components of the
Riemann tensor and its covariant derivatives in this frame have
the property that (i) all positive boost weight components (with
respect to $\ell$) are zero and (ii) all zero boost weight
components are constant.
\end{defn}
Note that \RC $\subset$ $CSI$ and \FC $\subset$ $CSI$. (There are
similar definitions for $CSI_{F,k}$ etc. \cite{epsilon}).

\begin{defn}[{\KC} spacetimes]
Finally, let us denote by \KC, those $CSI$ spacetimes that belong
to the Kundt class; the so-called Kundt $CSI$ spacetimes.
\end{defn}
We recall that a spacetime is Kundt on an open neighborhood if it
admits a null vector $\ell$ which is geodesic, non-expanding,
shear-free and non-twisting (which leads to constraints on the
Ricci rotation coefficients in that neighborhood; namely the
relevant Ricci rotation coefficients are zero). We note that if the
Ricci rotation coefficients are all constants then we have a locally
homogeneous spacetime.

\section{3D $CSI$ spacetimes}
\begin{thm}
Consider a 2-tensor $S_{\mu\nu}$ in $d$ dimensions. Then, if all
invariants up to the $d$th power of $S_{\mu\nu}$ are constants,
then the eigenvalues of the operator ${\sf S}\equiv
(S^{\mu}_{~\nu})$ are all constants.
\end{thm}
\begin{proof}
Consider the operator ${\sf S}\equiv (S^{\mu}_{~\nu})$ which maps vectors to vectors. Formally, the eigenvalues are determined by the equations $\det({\sf S}-\lambda {\sf 1})=0$. This can be expanded in the characteristic equation:
\[ \lambda^d+p_1[{\sf S}]\lambda^{d-1}+...+p_i[{\sf S}]\lambda^{d-i}+...+p_d[{\sf S}]=0\]
where $p_i[{\sf S}]$ are invariants of ${\sf S}$ of $i$th power. In particular,
\[ p_1[{\sf S}]=-S^{\mu}_{~\mu}, \quad p_2[{\sf S}]=\frac 12\left[(S^{\mu}_{~\mu})^2-S^{\nu}_{~\mu}S^{\mu}_{~\nu}\right],\cdots , \quad p_d[{\sf S}]=\pm \det[S^{\mu}_{~\nu}]. \]
By assumption, all coefficients of the eigenvalue equation are constant; hence, the solutions of the eigenvalue equation are all constants.
\end{proof}

 A consequence of
this is that for a $CSI_0$ spacetime all the eigenvalues of the
Ricci operator $R^{\mu}_{~\nu}$ are constants. In fact, we can do
better than this:
\begin{thm}
Consider a spacetime $\mathcal{M}$ and an open neighborhood
$U\subset\mathcal{M}$. If $\mathcal{M}$  has all constant zeroth
order curvature invariants and if the Segre type does not change
over $U$ then there exists a frame such that all the
components of the Ricci tensor are constants in $U$.
\label{Rconstant}\end{thm}
\begin{proof}
Petrov \cite{Petrov}.
\end{proof}

Indeed, it follows from Theorem 2.1 and the results of Petrov
\cite{Petrov} that this Theorem is true in general dimensions.
That is, on an open neighborhood in which the Segre type does not
change (and all Ricci invariants are constant) there exists a frame such that all of the components of the
Ricci tensor are constants and of a canonical form in any
dimension. This result is particularly powerful in 3D, where the
Riemann tensor is completely determined by the Ricci tensor.
However, the methods used here may not be directly generalized to
higher dimensions. For example, the frames in which the Ricci
tensor takes on it canonical Segre type and its special Weyl type
need not coincide.

In the 3D analysis below we shall explicitly use the canonical Segre forms for
the Ricci tensor in a frame in which the Ricci components are
constant.  A spacetime is said to be $k$-curvature homogeneous, denoted $CH$$_k$, if there exists a frame field in which the Riemann tensor and its covariant derivatives up to order $k$ are constant.  Consequently, we shall be considering the three dimensional $CH$$_0$ spacetimes. As usual, all of the
results are proven locally in an open neighborhood $U$ (in which
there exists a well-defined canonical frame for which the Segre
type does not change). Hence the results apply in neighborhoods
of all points except for a set of measure zero. In particular,
in 3D this means that in $U$, in which the Segre type does not change,
\[ CSI_0 \Leftrightarrow CH_0.\]
We will also show that in 3D
\[ CSI_2\Leftrightarrow CSI\]
Furthermore we will show that:
\begin{thm}
Assume that a 3D spacetime is locally $CSI$. Then, either:
\begin{enumerate}
\item{} the spacetime is locally homogeneous; \emph{or}
\item{} the spacetime is a \emph{Kundt spacetime} for which there exists a frame such that all curvature tensors have the following properties: (i) positive boost weight components all vanish; (ii) boost weight zero components are all constants.
\end{enumerate}
\end{thm}

This theorem applies locally in all open neighborhood in which
the Segre type does not change (see comments earlier).  Note that
the second part of this theorem states that all $CSI$ Kundt
spacetimes are also $CSI$$_F$ spacetimes; hence,
$CSI_K\subset CSI_F$. The proof of this theorem, done on a
case-by-case basis in terms of the Segre type,  is included in its
entirety in the following.

\subsection{$R^{\mu}_{~\nu}$ non-diagonal}
In what follows, we will choose a null frame such that the rotation one-forms are:
\beq
{\mbold\Omega}_{01}&=&A{\mbold\omega}^0+B{\mbold\omega}^{{1}}+C{\mbold\omega}^2, \label{eqs:Omega1} \\
{\mbold\Omega}_{02}&=&D{\mbold\omega}^0+E{\mbold\omega}^{{1}}+F{\mbold\omega}^2, \\
{\mbold\Omega}_{12}&=&G{\mbold\omega}^0+H{\mbold\omega}^{{1}}+I{\mbold\omega}^2. \label{eqs:Omega3}
\eeq The curvature 2-form can be determined by the Cartan
equations \beq {\bf R}^{\mu}_{~\nu}=\d
{\mbold\Omega}^{\mu}_{~\nu}+{\mbold\Omega}^{\mu}_{~\lambda}\wedge{\mbold\Omega}^{\lambda}_{~\nu},
\label{eq:R}\eeq where ${\bf
R}^{\mu}_{~\nu}=(1/2)R^{\mu}_{~\nu\alpha\beta}{\mbold\omega}^{\alpha}\wedge{\mbold\omega}^{\beta}$.
The Ricci tensor is now determined by contraction:
$R_{\mu\nu}=R^{\alpha}_{~\mu\alpha\nu}$.

In this null frame we will use the convention that the index 0 downstairs carries a negative boost weight while the index 1 downstairs carries positive boost weight.
For upstairs indices, the role is reversed.
(Note that this index  notation differs from that used in \cite{CSI}.) The relationship between
this formalism and notation and that of  \cite{CSI,Higher}
and  \cite{Hall} is discussed in  Appendix C.

\subsubsection{Segre type $\{21\}$}
Here we can choose a null-frame such that
\[ R_{01}=\lambda_1, \quad R_{00}=1, \quad R_{22}=\lambda_2\]
and we will assume $\lambda_1\neq \lambda_2$.

In this case we can define the operators:
\beq
{\sf P}_{1}=({\sf R}-\lambda_2{\sf 1}),\quad {\sf P}_2=({\sf R}-\lambda_1{\sf 1}).
\eeq
Now, we can define the projection operators
\beq
\perp_{1}&=&(\lambda_1-\lambda_2){\sf P}_1-{\sf P}_1{\sf P}_2, \\
\perp_{2}&=&{\sf P}_2^2,
\eeq
which projects onto the $01$ space and $2$ space respectively. Also useful is the operator $\perp_{3}\equiv {\sf P}_1{\sf P}_2$ which lowers a tensor 2 boost weights (the only nonzero component is $({\sf P}_1{\sf P}_2)^1_{~0}$).

Calculating $R^{\mu}_{~\nu;\lambda}$, and using the Bianchi identity, we get $I=0$,
$2B=F(\lambda_1-\lambda_2)$, and $H=-(G+E)(\lambda_1-\lambda_2)$.  Calculating
$S_{\alpha\beta\delta}=R_{\mu\nu;\lambda}(\perp_2)^{\mu}_{~\alpha}(\perp_1)^{\nu}_{~\beta}(\perp_1)^{\lambda}_{~\delta}$
gives the only non-zero components (omitting an irrelevant factor):  \beq S_{211}&=&
-H(\lambda_1-\lambda_2), \\ S_{210}&=&-G(\lambda_1-\lambda_2), \\
S_{201}&=&G(\lambda_1-\lambda_2), \\ S_{200}&=&-G-D(\lambda_1-\lambda_2).  \eeq
First, we
can form the invariant $S_{\alpha[\beta\gamma]}S^{\alpha[\beta\gamma]}$, and requiring
this to be constant yields $G$ constant.  Furthermore, using
$S_{\alpha\beta\gamma}(\perp_3)^{\beta}_{~\mu}(\perp_3)^{\gamma}_{~\nu}S^{\alpha\mu\nu}$
gives us $H$ constant.  The analysis splits into the cases in which $H$ is zero or not.

Assume $H\neq 0$ and constant. Then using $S_{\alpha\beta\gamma}S^{\alpha\beta\gamma}$
gives $D$ constant, while the Bianchi identities imply $E$ constant.
By suitable contractions with $S_{\alpha\beta\gamma}$, it can be immediately shown that all the remaining connection coefficients are constants. Hence, this is a locally homogeneous space. Furthermore,
$CSI$$_1$ implies $CH$$_1$ and also $CSI$.

Assume $H=0$ (i.e., Kundt).  This implies $G=-E$ (=constant).  Using the expressions for
the Ricci tensor (using eqs.  (\ref{eq:R})) we can show that $B$ has to satisfy:  \[
B_{,1}=-B^2, \] which also implies $\lambda_2=-2E^2$.  Using a boost we can choose $B=0$
($F$, however, will still remain non-zero in general).  This choice makes all the
connection coefficients of positive boost order vanish.  Let us now show that it is
$CSI$$_{F}$ using induction.  First, the above choice implies $CSI_{F,1}$. Therefore, assume
$CSI_{F,n}$.  Since there are no positive boost-weight connection coefficient, and from the
$CSI_{F,n}$ assumption, we have that $\left(\nabla^{(n+1)}R\right)_{1}=0$.  Then consider
$\left(\nabla^{(n+1)}R\right)_{0}$.  These components get contributions from
$\nabla\left(\nabla^{(n)}R\right)_0$, and possibly also
$\nabla\left(\nabla^{(n)}R\right)_{-1}$.  Now, the connection coefficients preserving the
boost weight are $C$, $G$ and $E$ of which $ G$ and $E$ are both constants.  Regarding
$C$, this comes from the connection one-forms
${\mbold\Omega}^0_{~0}=-{\mbold\Omega}^1_{~1}$.  Hence, due to the opposite sign we see
that $C$ does not contribute to $\left(\nabla^{(n+1)}R\right)_{0}$.  This implies that
$\nabla\left(\nabla^{(n)}R\right)_0$ obeys the $CSI_{F,n+1}$ criterion.  Finally, we need to
check boost weight
zero components of $\nabla\left(\nabla^{(n)}R\right)_{-1}$.  However, using the Bianchi
identities, and the identity \beq
[\nabla_{\mu},\nabla_{\nu}]T_{\alpha_1\cdots\alpha_i\cdots\alpha_m}=\sum_{i=1}^mR^{\lambda}_{~\alpha_i\mu\nu}T_{\alpha_1\cdots\lambda\cdots\alpha_m},
\eeq we see that $\nabla\left(\nabla^{(n)}R\right)_{-1}$ can only contribute with constant
components as well.  Hence, the space is $CSI_{F,n+1}$, and by induction, $CSI_F$.


\subsubsection{Segre type $\{(21)\}$} This is the case $\{21\}$ but with
$\lambda_2=\lambda_1$.  Here, the Bianchi identities give $H=0$, $I=2B$.  However,
all the components of $R_{\mu\nu;\lambda}$ have negative boost order.
All invariants of $R_{\mu\nu;\lambda}$ will therefore vanish.  Calculating the 2nd order
invariant $\Box R_{\mu\nu}\Box R^{\mu\nu}=96B^4$, where $\Box \equiv \nabla_{\mu}\nabla^{\mu}$,  we have that $B$ is thus a constant.  From the
Ricci tensor expressions, we have now $R_{11}=-6B^2=0$ which gives $B=I=0$.  This is
consequently a Kundt spacetime.

Furthermore, the Ricci equations give $C_{,1}=G_{,1}=0$, which allows us to boost away $C$ while keeping $B=0$. There is therefore no loss of generality to assume $C=0$. The remaining Ricci equations do now give additional differential equations for the remaining connection coefficients. Using a similar induction argument as for the Kundt spacetimes of Segre type $\{21\}$, these spacetimes are $CSI$$_F$.

\subsubsection{Segre type $\{3\}$ }
Here,  we can set
\[ R_{01}=R_{22}=\lambda, \quad R_{02}=1.\]
It is useful to define the projection operator (which has only boost weight -1 components):
\[ {\sf P}=({\sf R}-\lambda{\sf 1}), \]
for which ${\sf P}^2\neq 0$ (and has only boost weight -2 components), while ${\sf P}^3=0$.

Calculating $R_{\mu\nu;\lambda}$, and using the Bianchi identities, gives
\[ H=0, \quad B=2I, \quad G=-2E-C.\]
Now, the only boost weight zero components are proportional to $I$,
and all higher boost weight components vanish. Hence, by considering
$R_{\mu\nu;\lambda}R^{\mu\nu;\lambda}$ automatically gives $I$ is constant. The case now splits into whether $I$ is zero or not.

$I\neq 0$: By calculating the second order tensor $\Box R^{\mu}_{~\nu}$ (where $\Box=\nabla_{\mu}\nabla^{\mu}$), we get
the operator of the form
\[ \Box {\sf R}=\begin{bmatrix}
a & 0 & -3I^2 \\ c & a & b \\ b & -3I^2 & -2a
\end{bmatrix}.
\]
Here, $-3I^2$ corresponds to the boost weight +1 components, and $a$, $b$,
$c$ are defined to the  boost weight 0, -1, -2 components, respectively. We can now show that $a$, $b$, and $c$
are constants by calculating the invariants,
\[ \Tr\left[(\Box{\sf R})^2{\sf P}\right], \quad\Tr\left[(\Box{\sf R})^2\right], \quad \Tr\left[(\Box{\sf R})^3\right], \]
and requiring them to be constants. Hence, all components of $\Box R_{\mu\nu}$ are constants. Moreover, we can now use this operator and taking suitable contractions with $R_{\mu\nu;\lambda}$ to show that all connection coefficients are constants. This is therefore a locally homogeneous space.

$I=0$, Kundt case:  In this case all of the invariants of $R_{\mu\nu;\lambda}$ vanish
identically.  As can also be shown, all invariants of $R_{\mu\nu;\lambda\sigma}$ vanish
identically.  In fact, we can see that all invariants of all orders must vanish by
studying the connection coefficients and the Bianchi identities.  Note that $I=0$ implies
$B=0$, and since $H=0$ also, connection coefficients of positive boost weight are zero.
Again, using an induction argument, we can show that these Kundt spacetimes are also
$CSI$$_F$.

\subsubsection{Segre type $\{\bar{z}z1\}$}
This case is similar to the case $\{1,11\}$ below -- we just have to
consider complex projection operators.  At the end of the analysis, we get a similar result.
So in this case we have a locally homogeneous space and $CSI$$_1\Leftrightarrow $CH$_1$,
$CSI$$_1\Leftrightarrow$ $CSI$.

\subsection{$R^{\mu}_{~\nu}=\diag(\lambda_0,\lambda_1,\lambda_2)$}
\subsubsection{Segre type $\{1,11\}$: Eigenvalues all distinct}
Now, we can define the projection operators:
\beq
{\sf P}_0&=&({\sf R}-\lambda_1{\sf 1})({\sf R}-\lambda_2{\sf 1})\\
{\sf P}_1&=&({\sf R}-\lambda_0{\sf 1})({\sf R}-\lambda_2{\sf 1})\\
{\sf P}_2&=&({\sf R}-\lambda_0{\sf 1})({\sf R}-\lambda_1{\sf 1}).
\eeq
We note that ${\sf P}_0{\sf P}_1=0$ etc., so that these projection operators project orthogonally onto the respective eigenvectors. Furthermore, ${\sf P}_0{\bf v}_{(0)}=(\lambda_0-\lambda_1)(\lambda_0-\lambda_2){\bf v}_{(0)}$, etc. It is therefore essential that all of the eigenvalues are distinct. Note also that the projection operators are made out of curvature tensors and their invariants; hence, they are curvature tensors themselves.

We can now consider curvature tensors of the type:
\[ R(ijk)_{\alpha\beta\delta}\equiv R_{\mu\nu;\lambda}({\sf P}_i)^{\mu}_{~\alpha}({\sf P}_j)^{\nu}_{~\beta}({\sf P}_k)^{\lambda}_{~\delta} \]
for which the invariant $R(ijk)_{\alpha\beta\delta}R(ijk)^{\alpha\beta\delta}$
is essentially the square of the component $R_{ij;k}$ (up to a constant factor). Requiring that all such
are constants implies that with respect to the aforementioned frame, all connection coefficients are constants. This, in turn, implies that the spacetime is a locally homogeneous spacetime.
So in this case: $CSI$$_1\Leftrightarrow$ $CH$$_1$, $CSI$$_1\Leftrightarrow CSI$.

\subsubsection{Segre type $\{(1,1)1\}$: $\lambda_0=\lambda_1$}
This case is similar to $\{21\}$. Using a null-frame, the Bianchi identities give
$I=F=0$ and $G=-E$. Similarly, defining $S_{\mu\nu\lambda}$, we obtain
\beq
S_{211}&=& -H(\lambda_1-\lambda_2), \\
S_{210}&=&-G(\lambda_1-\lambda_2), \\
S_{201}&=&G(\lambda_1-\lambda_2), \\
S_{200}&=&-D(\lambda_1-\lambda_2),
\eeq
which again implies  that $G$ is constant. Furthermore, $DH$ is constant. We can
now boost so that $D$ is constant. This, in turn, implies that  $H$ constant and the spacetime is $CH$$_1$.

$H\neq 0$: From the equations for the Ricci tensor we get $A=B=C=0$. This implies a locally homogeneous space.

$H=0$, Kundt case:  This further splits into 2 cases:  \begin{enumerate} \item{}
$D\neq 0$:  The equations for the Ricci tensor imply $B=C=0$ and $\lambda_2=-2E^2$.  For
$A$ we get the differential equations:  \beq A_{,1}=\lambda_1, \quad A_{,2}=EA.  \eeq
These are eqs. (51)-(53) in \cite{Bueken} and correspond to an inhomogeneous spacetime as
long as $A$ is non-constant.  We can see that this is not $CH$$_2$ by considering \[
R_{02;00}=-2DA(\lambda_1-\lambda_2).\] However, using an induction argument, as before,
we can show that this Kundt spacetime is $CSI$$_F$.  \item{} $D=0$:  Here, we can use a
boost to set $B=0$.  This, in turn, implies that $C_{,1}=0$.  Hence, we can simultaneously solve
\[ \rho_{,1}=0,\quad \rho_{,2}=-C, \] to also boost away $C$.  We are now left with \beq
A_{,1}=\lambda_1, \quad A_{,2}=EA.  \eeq
However, as can easily be checked, the isotropy group
of $R_{\mu\nu;\lambda}$ is 1-dimensional (as is the isotropy subgroup for $R_{\mu\nu}$),
and by Singer's theorem \cite{Singer} this is a locally homogeneous space.  These are therefore
trivially $CSI$$_F$.  \end{enumerate}

\subsubsection{Segre type $\{1,(11)\}$: $\lambda_1=\lambda_2$}
In this case we define the projection operators
\beq
{\sf P}_t&=&({\sf R}-\lambda_1{\sf 1})\\
{\sf P}_s&=&({\sf R}-\lambda_0{\sf 1}),
\eeq
which project onto the time-like, and space-like eigendirections, respectively.

Assuming an \emph{orthonormal frame} with connection one-forms as earlier
(with 0 being time-like), the Bianchi identities
imply $A=D=0$ and $B=-F$. The remaining components of $R_{\mu\nu;\lambda}$ are:
\beq
(R^a_{~0;b})=\begin{bmatrix}
B & C \\ F & -B
\end{bmatrix}, \qquad a,b=1,2.
\eeq
We can always use a $U(1)$ rotation to set $B=0$. Then, the
antisymmetric/symmetric parts give $C$ and $F$ constants. Hence, this is always
$CH$$_1$ and it follows that this case is therefore locally homogeneous \cite{Bueken}.

\subsubsection{Segre type $\{(1,11)\}$: $\lambda_0=\lambda_1=\lambda_2$}
By calculation $R_{\mu\nu;\lambda}=0$, so this is the maximally
symmetric case. This case is automatically $CSI$ since all the
higher-order curvature invariants vanish identically. In
particular, this implies that it is $CH$$_k$ for all $k$ and hence
locally homogeneous. In fact, in 3D there are only three
possibilities, namely de Sitter (dS$_3$), Anti-de Sitter (AdS$_3$) and Minkowski
space, all of which are also Kundt spacetimes. Trivially, they are
also $CSI$$_F$ spacetimes.

\section{Locally homogeneous and Kundt $CSI$ spacetimes in 3D}
\subsection{Locally homogeneous spaces}
The 3D locally homogeneous Lorentzian spacetimes were recently classified by Calvaruso
\cite{Calvaruso}. The main theorem is as follows:
\begin{thm}
Let $(\mathcal{M},g)$ be a three-dimensional Lorentzian manifold. The following conditions are equivalent:
\begin{enumerate}
\item{} $(\mathcal{M},g)$ is curvature homogeneous up to order two;
\item{} $(\mathcal{M},g)$ is locally homogeneous;
\item{} $(\mathcal{M},g)$ is either locally symmetric, or locally isometric to a Lie group equipped with a left-invariant Lorentzian metric.
\end{enumerate}
\end{thm}

This theorem is extremely powerful and gives all the locally homogeneous spacetimes. Trivially, they are also $CSI$ spacetimes and determines the set $H\subset CSI$. To get the actual metrics one needs to determine the possible metrics satisfying the theorem. This was done in \cite{Calvaruso}.

\subsection{3D Kundt $CSI$ spacetimes} We have now established that an
inhomogeneous 3D $CSI$ spacetime must be $CSI$$_F$ and $CSI$$_K$.  In
\cite{CSI} it was shown that such a spacetime can be written:
\beq
\d s^2=2\d u\left[\d v +H(v,u,x)\d u+W_x(v,u,x)\d x\right] +\d
x^2, \label{Kundt}\eeq where \beq
W_x(v,u,x) &=& vW^{(1)}_x(u,x)+W^{(0)}_x(u,x), \\
H(v,u,x) &=& \frac{v^2}{8}\left[ 4\sigma+\left(W^{(1)}_x\right)^2\right]+vH^{(1)}(u,x)+H^{(0)}(u,x),
\eeq
and $\sigma$ is a constant. Note that, in general, this frame is not the same frame
that was considered earlier.

From the $CSI$$_0$ and $CSI$$_1$ criteria, the function $W_x^{(1)}$ fulfills the equations:
\beq
\partial_x W^{(1)}_{x}-\frac 12\left(W^{(1)}_x\right)^2&=&s, \label{eq:CSI0} \\
\left(2\sigma-s\right)W^{(1)}_x &=&2\alpha, \label{eq:CSI1}
\eeq
where $s$ and $\alpha$ are constants.

We can see that the Kundt $CSI$ spacetimes split into two cases,
according to whether $2\sigma-s$ is zero or not.

\subsubsection{Segre types $\{21\} $ and $\{(1,1)1\}$: $s\neq 2\sigma$} We see that in this case the only solutions to eq.(\ref{eq:CSI1}) are
\[ W^{(1)}_x=2r, \]
where $r$ is a constant. From (\ref{eq:CSI0}) this implies $s=-2r^2$.

All of these cases are therefore:
\beq
W_x(v,u,x) &=& 2rv+W^{(0)}_x(u,x), \\
H(v,u,x) &=& {v^2}\tilde{\sigma}+vH^{(1)}(u,x)+H^{(0)}(u,x),
\eeq
where $\tilde{\sigma}=(\sigma+r^2)/2$.

\subsubsection{Segre types $\{3\}$, $\{(21)\}$ and $\{(1,11)\}$: $s=2\sigma$}
In this case eq.(\ref{eq:CSI1}) is identically satisfied. Solving eq. (\ref{eq:CSI0})
gives us the following cases (where the $u$-dependence has been eliminated using a coordinate transformation):
\begin{enumerate}
\item{} $\sigma>0$:
$W^{(1)}_x=2\sqrt{\sigma}\tan\left(\sqrt{\sigma}x\right)$. \item{}
$\sigma=0$: $W^{(1)}_x=\frac{2\epsilon}{x}$, where $\epsilon=0,1$.
This is the $VSI$ case. \item{} $\sigma<0$:
$W^{(1)}_x=-2\sqrt{|\sigma|}\tanh\left(\sqrt{|\sigma|}x\right)$.
\item{} $\sigma<0$:
$W^{(1)}_x=2\sqrt{|\sigma|}\coth\left(\sqrt{|\sigma|}x\right)$.
\item{} $\sigma<0$: $W^{(1)}_x=2\sqrt{|\sigma|}$.
\end{enumerate}
All of these metrics were given in \cite{CSI}. The case with $\sigma>0$
has the same invariants as dS$_3$, while the cases with $\sigma<0$ have the same invariants as AdS$_3$. The case $\sigma=0$ has all vanishing curvature invariants.

This is therefore an exhaustive list of Kundt $CSI$ spacetimes in 3D.  We are now in a situation
to state:
\begin{thm} Consider a 3D $CSI$ spacetime, $(\mathcal{M},g)$.  Then there exists a
locally homogeneous space, $(\widetilde{\mathcal{M}},\widetilde{g})$ having the same curvature
invariants as $(\mathcal{M},g)$.  \end{thm}
\begin{proof} Consider a 3D $CSI$ spacetime
$(\mathcal{M},g)$.  This spacetime has to be either locally homogeneous, in which case we can
set $(\mathcal{M},g)=(\widetilde{\mathcal{M}},\widetilde{g})$ and the theorem is trivially
satisfied, or Kundt.  Assume therefore the spacetime is Kundt.  All possible Kundt spacetimes
are listed above.  If $s\neq 2\sigma$, $W^{(1)}_x=2r$ is constant.  In this case we can let
$(\widetilde{\mathcal{M}},\widetilde{g})$ be the Kundt spacetime where $W^{(1)}=2r$, $\widetilde{\sigma}=(\sigma+r^2)/2$, $W^{(0)}=H^{(1)}=H^{(0)}=0$.  This can be seen to be a locally homogeneous space by choosing the
left-invariant frame ${\mbold\omega}^{0}=v\d u$, ${\mbold\omega}^1=\d v/v+\widetilde{\sigma}v\d u+2r\d x$ and
${\mbold\omega}^2=\d x$.  Finally, if $s=2\sigma$, then we choose
$(\widetilde{\mathcal{M}},\widetilde{g})$ to be de Sitter space, Minkowski space, and Anti-de
Sitter space for $\sigma>0$, $\sigma=0$ and $\sigma<0$, respectively.  \end{proof}

\section{Discussion}

In this paper we have explicitly found all 3D $CSI$ metrics. In
particular, we have proven the various $CSI$ conjectures in 3D,
shown that for every $CSI$ with particular constant invariants
there is a locally homogeneous spacetime with precisely the same
constant invariants, and demonstrated that for $CSI$ spacetimes
that are not locally homogeneous, the Ricci type is $II$ or less
 and that all boost weight zero terms are constant.

In more detail, in 3D we have proven that a spacetime is locally
$CSI$ if and only if there exists a null frame in which the Riemann
tensor and its derivatives are either constant, in which case we
have a locally homogeneous space, or are of boost order zero with
constant boost weight zero components at each order so that the
Riemann tensor is of type II or less (i.e., we have proven the
${CSI_F}$ conjecture). We have also proven that if a 3D spacetime
is locally {$CSI$}, then the spacetime is either locally
homogeneous or belongs to the Kundt $CSI$ class (the ${CSI_K}$
conjecture) Finally, we have explicitly demonstrated the validity
of the ${CSI_R}$ conjecture in 3D by finding all such spacetimes
and showing how they are constructed from locally homogeneous
spaces and $VSI$ spacetimes (by means of fibering and warping).

In the analysis we have used the canonical Segre forms for
the Ricci tensor in a frame in which the Ricci components are all
constant \cite{Milson}. We used these canonical forms to
prove a number of results in 3D \cite{Hall}. We classified
different cases in terms of  their Segre type: the Segre type is
more refined than the Ricci type and in 3D the Riemann tensor is
completely determined by the Ricci tensor. As usual, all of the
results are proven locally in an open neighborhood $U$ (in which
there exists a well-defined canonical frame for which the Segre
type does not change). Hence the results apply in neighborhoods
of all points except for a set of measure zero.

We could also attempt to prove the results in terms of Ricci
types. We can easily write down the classification of the Ricci
tensor according to boost weights in a chosen frame \cite{class}. We could then
seek an alternative proof of the main results, particularly in the
crucial case of Ricci type II. However, in an open neighborhood
in which the Ricci type stays the same, the Segre type can change.  In Table \ref{table1} we give the relationship between Ricci type and Segre type.
Therefore, unlike in the case of Segre types, it may not be possible in general
to find a frame in which all components of Ricci tensor are constants
in the whole neighborhood. In this alternative proof we would
need, for each Ricci type, to simplify (by choice of frame) the
higher boost weight components of the Ricci tensor and the Ricci
rotation coefficients, and then proceed to utilize the constant
scalar (differential) invariants to prove the relevant results.

Note that the Theorem \ref{Rconstant} does not apply to points at which the Segre type changes.  Consider
therefore a point $p\in \mathcal{M}$ for which no neighborhood exists such that the
Segre type is the same.  Let us assume that $\mathcal{M}$ is a 3D $CSI$$_0$ spacetime.
Since the eigenvalues are constant, at $p$ we may have one of the following degeneracies:
\beq &&\{21\} \rightarrow \{(1,1)1\}, \nonumber \\ &&\{3\} \rightarrow \{(21)\}
\rightarrow \{(1,11)\}.  \eeq Clearly, if the Ricci components are constants, then the
Segre type cannot change, so the non-changing Segre type criterion in Theorem
\ref{Rconstant} is essential.  It is also essential to consider Segre type, and not Ricci
type, since both $\{3\}$ and $\{(21)\}$ are of Ricci type II.
Let us consider a simple example where such a degeneracy occurs.
The following is, in general, a Ricci type II metric:
\[
\d s^2=2\d u\left(\d v+\left[q(u)x^2v+H^{(0)}(u,x)\right]\d u+2pv\d x\right)+\d x^2.
\]
For $q(u)x\neq 0$ this is of Segre type $\{3\}$, however, along the line $x=0$
this degenerates to Segre type $\{(21)\}$.

A classification of 3D curvature homogeneous Lorentzian manifolds was carried out
in \cite{Bueken}.  It was shown that in all Segre types, $CH$$_2$ implies local homogeneity and
the only proper $CH$$_1$ (not locally homogeneous) spacetimes occur in the degenerate Segre
types $\{(21)\}$ or $\{(1,1)1\}$.  In both cases the classes of solutions have been
determined and shown to be parameterized by one function of one variable \cite{Bueken}.  More
recently, a study of four dimensional curvature homogeneous spacetimes has shown that
$CH$$_3$ implies local homogeneity \cite{Mp}; in addition, there exists a class of proper $CH$$_2$
spacetimes of Petrov type N, Plebanski-Petrov type N with a negative cosmological constant.  The
class of proper $CH$$_2$ spacetimes has been explicitly determined and shown to depend
on one function of one variable \cite{Mp}.  In contrast, three and four dimensional Riemannian manifolds
that are $CH$$_1$ are necessarily locally homogeneous \cite{sekigawa,ssv}.

In the case of four dimensional $CSI$ spacetimes, previously it has
been argued  \cite{CSI} that the spacetime is  either of Petrov (Weyl)
type I and Plebanski-Petrov (Ricci) type I and that it is plausible that
all such spacetimes are locally homogeneous, or that  the spacetime is
of Petrov type II and Plebanski-Petrov type II and it follows that
all boost weight zero terms are necessarily constant and if the
spacetime is not locally homogeneous a number of further
restrictive conditions apply (that support the validity of the
$CSI$ conjectures in four dimensions). Our aim is to next prove
the $CSI$ conjectures in four dimensions by considering each Petrov-type
and Plebanski-Petrov type (or Segre type) separately, by
investigating a number of appropriate differential scalar
invariants, and exploiting the insights gained in the present
work. Ultimately, our goal is to study $CSI$ spacetimes in
arbitrary higher dimensions, classified according to their Weyl
type and Ricci or Segre type.

\appendix

\section{Boosting the connection coefficients}
Consider a null-frame and assume the rotation one-forms are:
\beq
{\mbold\Omega}_{01}&=&A{\mbold\omega}^0+B{\mbold\omega}^1+C{\mbold\omega}^2, \\
{\mbold\Omega}_{02}&=&D{\mbold\omega}^0+E{\mbold\omega}^1+F{\mbold\omega}^2, \\
{\mbold\Omega}_{12}&=&G{\mbold\omega}^0+H{\mbold\omega}^1+I{\mbold\omega}^2.
\eeq
Then, under the following boost:
\beq {\widetilde{\mbold\omega}}^0=e^\rho{\mbold\omega}^0, \quad  {\widetilde{\mbold\omega}}^1=e^{-\rho}{\mbold\omega}^1, \quad  {\widetilde{\mbold\omega}}^2={\mbold\omega}^2, \label{boostcoframe}
\eeq
the connection coefficients will change according to
\beq
&&\widetilde{A}=e^{-\rho}(\rho_{,0}+A), \quad \widetilde{B}=e^{\rho}(\rho_{,1}+B), \quad \widetilde{C}=\rho_{,2}+C,\nonumber \\
&& \widetilde{D}=e^{-2\rho}D, \quad \widetilde{E}=E, \quad \widetilde{F}=e^{-\rho}F, \nonumber \\
&& \widetilde{G}=G, \quad \widetilde{H}=e^{2\rho}H, \quad \widetilde{I}=e^{\rho}I.
\eeq

\noindent The boost weight decomposition of the Ricci tenor is
\begin{eqnarray}
R_{ab} & = & R_{11}n_{a}n_{b}+2R_{12}n_{(a}m_{b)}+2R_{10}n_{(a}\ell_{b)}+R_{22}m_{a}m_{b} \nonumber\\
& & \mbox{} \hspace{6.4cm}+2R_{02}\ell_{(a}m_{b)} + R_{00}\ell_{a}\ell_{b} \label{Riccidecomp}
\end{eqnarray}
where we have identified ${\mbold\ell}={\mbold\omega}^0$, ${\mbold n}={\mbold\omega}^1$ and ${\mbold m}={\mbold\omega}^2$.  A frame component $T$ of a tensor ${\mbold T}$ has boost weight $b$ if subject to a boost of (\ref{boostcoframe}) transforms as $\widetilde{T}=e^{b\rho}T$.  It follows from (\ref{Riccidecomp}) that the Ricci frame scalars $\{ R_{11}, R_{12},R_{10}, R_{22}, R_{02}, R_{00}\}$ have boost weights $\{+2,+1,0,0,-1,-2\}$ respectively.  An algebraic classification of tensors for arbitrary dimensions has been given in \cite{Milson} and \cite{class}, this can be used to classify the Ricci tensor.  The Ricci tensor type is determined by the existence of a frame, in a neighborhood, for which certain boost weights of the Ricci frame scalars vanish, these are summarized in Table \ref{Riccitypetable}.
\begin{table}
\caption{Ricci type in terms of the vanishing of Ricci frame scalars.}\label{Riccitypetable}
\centering
\begin{tabular}{cl}
Ricci type & Ricci scalars \\
\hline
I & $R_{11}=0$ \\
II & $R_{11}=R_{12}=0$ \\
D & $R_{11}=R_{12}=R_{02}=R_{00}=0$ \\
III & $R_{11}=R_{12}=R_{10}=R_{22}=R_{00}=0$ \\
N & $R_{11}=R_{12}=R_{10}=R_{22}=R_{02}=0$ \\
O & $R_{\alpha\beta}=0$
\end{tabular}
\end{table}

\section{Kundt Spacetimes}
Consider the null vector $k_{\mu}{\mbold\omega}^{\mu}\equiv{\mbold\omega}^0$.
Then the components of the covariant derivative are:
\beq
&& k_{0;0}=A, \quad k_{0;1}=B, \quad k_{0;2}=C, \nonumber \\
&& k_{1;0}=k_{1;1}=k_{1;2}=0, \nonumber \\
&& k_{2;0}=-G, \quad k_{2;1}=-H, \quad k_{2;2}=-I.
\eeq
 Note that the boost weight +1 component is $-H$, while the boost weight 0 components are $-I$ and $-B$. Hence, a sufficient criterion for the spacetime to possess a geodesic, expansion-free, shear-free, twist-free null-vector (i.e., $k^{\nu}k_{\mu;\nu}=k^{\mu}_{~;\mu}=k_{(\mu;\nu)}k^{\mu;\nu}=k_{[\mu;\nu]}k^{\mu;\nu}=k_{\mu}k^{\mu}=0$)
is that there exists a frame such that $H=I=B=0$. In fact, using a boost above, $H=I=0$, is sufficient.

Any 3D Kundt metric can be written in the from of eq. (\ref{Kundt}). For this 'Kundt frame' we have
\beq
{\mbold\omega}^0=\d u, \quad {\mbold\omega}^1=\d v+{H}\d v+W\d x, \quad {\mbold\omega}^2=\d x.
\eeq
For this frame the connection coefficients are:
\beq
&&A=H_{,v}, \quad B=0, \quad C=\frac 12W_{x,v}, \nonumber \\
&&D=H_{,x}-H_{,v}W_x+HW_{x,v}-W_{x,u}, \quad E=-\frac 12W_{x,v}, \quad F=0, \nonumber \\
&&G=-\frac 12W_{x,v}, \quad H=0, \quad I=0.
\eeq

\begin{table}
\caption{Ricci type to Segre type conversion scheme. }\label{table1}
\centering
\begin{tabular}{|c|c|c|}
\hline
Segre type &  & Ricci type \\
\hline\hline
$\{1,11\}$ & & $I$ \\
\hline
$\{1,(11)\}$ & & $I$ \\
\hline
$\{(1,1)1\}$ & & $D$ \\
\hline
$\{(1,11)\} $ & $\lambda_1\neq 0$ & $D$ \\
              & $\lambda_1=0$ & $O$ \\
\hline
$\{1\bar{z}z\}$ && $I$ \\
 \hline
$\{21\}$ & & $II$ \\
\hline
$\{(21)\}$ & $\lambda_1\neq 0$& $II$ \\
           & $\lambda_1= 0$ & $N$ \\
\hline
$\{3\}$ & $\lambda_1\neq 0$& $II$ \\
           & $\lambda_1= 0$ & $III$ \\
\hline
\end{tabular}
\end{table}

\section{Comparisons with other work}

In the analysis we have chosen the frame ${\mbold e}_{\alpha}=\{{\mbold n}, {\mbold \ell},
{\mbold m}\}$ with inner product

\begin{equation}
\eta_{\alpha\beta}=\left[
\begin{array}{ccc}
0 & 1 & 0 \\
1 & 0 & 0 \\
0 & 0 & 1
\end{array}
\right] \, .
\end{equation}
We may relate the connection coefficients of (\ref{eqs:Omega1})--(\ref{eqs:Omega3}) with the
Ricci rotation components $L_{\alpha\beta}$, defined by \cite{Higher,bianchi}:

\begin{equation}
\ell_{a;b}=L_{11}\ell_{a}\ell_{b}+L_{10}\ell_{a}n_{b}+L_{1i}\ell_{a}m^{i}_{\ b}+L_{i1}m^{i}_{\ a}\ell_{b} + L_{i0}m^{i}_{\ a}n_{b}+L_{ij}m^{i}_{\ a}m^{j}_{\ b} \, .
\end{equation}
In the 3D case $i,j=2$, and we straightforwardly find that
\begin{eqnarray}
L_{11}=A & L_{10}=B & L_{12}=C \\
L_{21}=-G & L_{20}=-H & L_{22}=-I \, .
\end{eqnarray}
Therefore, $\ell$ is geodesic if $L_{i0}=-H=0$,
affinely parameterized if $L_{10}=B=0$ and expansion-free if $L_{ij}=-I=0$
(it is automatically shear and twist free).

In other work, a real Newman-Penrose frame has been used to study 3D spacetimes \cite{Hall}.
In order to relate the notation used here with that in \cite{Hall},
we first note their choice of null frame is ${\mbold z}_{\alpha}=\{{\mbold z}_{o}={\mbold m},
{\mbold z}_{+}={\mbold \ell}, {\mbold z}_{-}={\mbold n}\}$.
Therefore, the indices used here and those used in \cite{Hall} are
related by $o\leftrightarrow 2$, $+\leftrightarrow 1$ and $-\leftrightarrow 0$.  Moreover, they define the Ricci rotation coefficients by
\begin{equation}
{\mbold \gamma}_{\mu\nu\rho}={\mbold z}_{\mu i;j}{\mbold z}_{\nu}^{\ \ i}{\mbold z}_{\rho}^{\ \ j}  \, ,
\end{equation}
so that $\Omega_{\mu\nu\rho}=-\gamma_{\mu\nu\rho}$, where
$\Omega_{\mu\nu\rho}$ are connection coefficients defined in
(\ref{eqs:Omega1})--(\ref{eqs:Omega3}).
Therefore, we have the following relations (see (6) of \cite{Hall})

\begin{eqnarray}
\tilde{\kappa}=-F & \kappa=-I & \varepsilon=C \\
\tilde{\rho}=-E & \sigma=-H & -\tilde{\tau}=B \\
\tilde{\sigma}=-D & \rho=-G & \tau=A \, .
\end{eqnarray}


\section*{Acknowledgments}
This work was supported by the
Natural Sciences and Engineering Research Council of Canada and
AARMS (SH).

\end{document}